# Azimuthal and polar anchoring energies of aligning layers structured by nonlinear laser lithography

I. Pavlov[a,b], O. Candemir[b], A. Rybak[a], A. Dobrovolskiy[c], V. Kadan[a], I. Blonskyi[a], P. Lytvyn[d], A. Korchovyi[d], P. Tytarenko[e], Z. Kazantseva[f] and I. Gvozdovskyy[g]*

[a] *Department of Photon Processes, Institute of Physics of the National Academy of Sciences of Ukraine, Kyiv, Ukraine;* [b] *Department of Physics, Middle East Technical University, Ankara, Turkey;* [c] *Department of Gas Electronics, Institute of Physics of the National Academy of Sciences of Ukraine, Kyiv, Ukraine;* [d] *Department of Structural and Elemental Analysis of Materials ans Systems, V.E. Lashkaryov Institute of Semiconductor Physics of the National Academy of Sciences of Ukraine, Kyiv, Ukraine;* [e] *Department of Optoelectronic Llightgenerating Devices and Systems, V.E. Lashkaryov Institute of Semiconductor Physics of the National Academy of Sciences of Ukraine, Kyiv, Ukraine;* [f] *Department of Electrical and Galvanomagnetic Properties of Semiconductors, V.E. Lashkaryov Institute of Semiconductor Physics of the National Academy of Sciences of Ukraine, Kyiv, Ukraine;* [g] *Department of Optical Quantum Electronics, Institute of Physics of the National Academy of Sciences of Ukraine, Kyiv, Ukraine*

Institute of Physics of the National Academy of Sciences of Ukraine, Prospekt Nauki 46, Kyiv-28, 03028, Ukraine, telephone number: +380 44 5250862, *E-mail: igvozd@gmail.com

# Azimuthal polar anchoring energies of aligning layers structured by nonlinear laser lithography

In spite of the fact that there are different techniques in the creation of the high-quality liquid crystals (LCs) alignment by means of various surfaces, the azimuthal and polar anchoring energies as well as the pre-tilt angle are important parameters to all of them. Here, the modified by a certain manner aligning layers, previously formed by nonlinear laser lithography (NLL), having high-quality nano-periodic grooves on Ti surfaces, recently proposed for LC alignment was studied. The change of the scanning speed of NLL in the process of nano-structured Ti surfaces and their further modification by means of ITO-coating, and deposition of polyimide film has enabled different aligning layers, whose main characteristics, namely azimuthal and polar anchoring energies, were measured. For the modified aligning layers, the dependencies of the twist and pre-tilt angles for LC cells filled by nematic E7 ($\Delta\varepsilon > 0$) and MLC-6609 ($\Delta\varepsilon < 0$) were obtained. Also the contact angle for droplets of isotropic liquid (glycerol), and nematic LCs was measured for the various values of the scanning speed during the laser processing.

**Introduction**

It is relevant to note that today's industrial production of the liquid crystal displays (LCDs) is very intense. The alignment of LCs is an important stage of the production of various LC devices. It is known that the most commonly encountered method of the orientation of LCs, used in the LCD technology is the rubbing technique. However, the studies of various aligning layers and methods of their creation, despite the fact that the production of LCDs are continued, remains as before a very important task for both the application and the viewpoint of science.

There are several different causes of the nature of alignment of the LCs, such as the creation of the rough-finished surface during rubbing technique, the creation of the charges at the surface during rubbing, photoalignment or ion/plasma techniques [1-9]. However, the main cause of alignment properties of the surface there is the creation of the anisotropic distribution of the roughness, charges and cross-linking/destruction of the polymer [1-20]. Each of these methods of the treatment of surface has their own advantages and drawbacks. For example, the rubbing technique widely used in LCD technology has some shortcomings, such as accumulation of both the state charges and dust particles [1]. The photoalignment technique for the first time was introduced by Ichimura [2] for controlling alignment of LC in zenithal plane. Thereafter the homogeneous photoalignment of LCs in an azimuthal plane was simultaneously discovered by groups of W. Gibbons [3], M. Schadt [4] and Yu. Reznikov [5,6]. Photoalignment is a really alternative method to the rubbing technique, because this technique leads to significant simplification of the change of the orientational order of photoproducts under polarized light. However, main shortcomings of photoaligning technique are low switching speed of LC devices and gradually deteriorating of the polymer layers. The usage of a plasma beam as a perspective method of processing aligning layers for the homogeneous planar and tilted orientation of LCs was studied in [7,8]. However, the ion/plasma beam processing is complicated and expensive, having some shortcomings such as accumulation of static charges, wrack of the polymer layer or deterioration of their properties. In addition, e-beam lithography [9], atomic force microscopy (AFM) nano-rubbing [10,11], nano-imprint lithography [12,13] and photolithography [14] were also used to obtain the LC alignment. However, these methods show a very low throughput with a very small treated areas and limited period of nano-grooves. It should be also noted that a fast and high-throughput crack-induced grooving method [15] was recently proposed. The alignment of the nematic on a pre-rubbed polyimide film processed by laser-induced periodic surface structuring (LIPSS) [16] using Nd:YAG laser (third harmonic frequency λ = 355 nm) was studied [17]. Recently, intense researches of laser processing with femtosecond pulse on purpose to process of different materials enabled the high-quality periodic nano-grooves of a titanium (Ti) [18-20] and ITO [21] layer, which were proposed as a simple technique of the LC alignment by creating a large areas micro or high-quality nano-grooves.

All above-mentioned techniques for alignment of LCs allow in certain ways to carry out more or less the effective control of the main parameters, such as easy orientation axis, pre-tilt angle and anchoring energy (AE). Hence, the interest of studying in the surface properties of LCs is always great [1]. A certain surface orientation of the director $\vec{n}$ there exists owing to an excess of the free energy per unit surface area. In an equilibrium state when free energy is minimal, the director $\vec{n}$ is aligned along an easy axis defined by the two angles, namely twist angle $\varphi$ in the plane of the aligning layer (so-called azimuthal angle) and pre-tilt angle $\theta_p$ deposited in the plane, which is perpendicular to the plane of aligning layer (so-called polar angle), as shown in Figure 1.

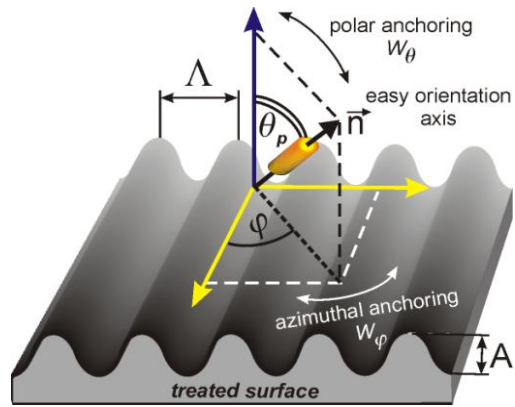

Figure 1. Schematic presentation of director $\vec{n}$ in the bulk of LC, which is bordered on the nano-structured aligning layer.

One of the most important characteristics of the aligning layer certainly is AE, namely azimuthal anchoring energy (AAE $W_\varphi$) and polar anchoring energy (PAE $W_\theta$) which will be described in this manuscript. By knowing the AE of aligning surfaces and having a way for controllable changing it, we can use these surfaces for different applications. For example, surfaces possessing strong AAE $W_\varphi$ is needed in order to receive the strong homogeneous alignment of LCs and can be used for storage some information. In contrary, to obtain the better characteristics in dynamic processes, namely the low-voltage switching of LCs from planar (or tilted) to homeotropic alignment (vertical, VA) or vice versa from planar to VA will take place for surfaces possessing weak AAE $W_\varphi$ and strong PAE $W_\theta$. It is well known that there are several calculation methods for AE values of aligning surfaces [22-37].

It is obvious that the orientation of the LCs depends on the AE between molecules of LC and aligning layer. Furthermore, any surfaces, including the aligning layers, are also characterised by surface tension $\gamma_S$ [38]. Consequently, some surfaces possess a hydrophobic property while other surfaces are characterized by a hydrophilic feature. Owing to this fact there might be different contributions of LC molecules orientation on the aligning surface in bulk of LC cell (Figure 1) [39].

According to empirical Creagh-Kmetz's rule [38,39] the homeotropic alignment of nematic LC is obtained at $\gamma_S < \gamma_{LC}$ while the planar alignment occurs at $\gamma_{LC} < \gamma_S$ ($\gamma_S$ is the surface tension of the substrate and $\gamma_{LC}$ is the surface tension of LC). Recently it was shown that on the one hand, there has been certain exposure dependence of the contact angle of the nematic droplet E7 at the photosensitive PVCN-F surface [40] and on the other hand, the irradiation time dependence of the AAE $W_\varphi$ of PVCN-F surfaces was found [36]. It is important to note that the influence of the laser power on both the adhesion, surface free energy graphene treated by nonlinear laser lithography (NLL) [41], and contact angle of the water droplets deposited on the various treated graphene surfaces was studied [42].

Recently, we studied the nano-structured Ti layer by means of the NLL as aligning layer for the nematic LCs orientation [19,20]. For pure nano-structured Ti layers and Ti layers coated by polyimide ODAPI film the AAE $W_\varphi$ was determined by using the method of combined twist LC cell. It was shown that the value of the AAE $W_\varphi$ of aligning layer depends on various parameters of the NLL, namely laser pulse fluence (LPF, $J$) and scanning speed $v$ [19,20]. It was therefore logical the next stage study for characterisation of nano-structured Ti layer as an aligning surface of LCs in terms of PAE $W_\theta$, depending on various parameters of the laser processing. However, currently, there exist a few well-known methods based on the usage of the alternating electric field [22-34] to estimate the PAE $W_\theta$ of the aligning layer. Therefore, LC cells besides of the aligning layer, additionally posses an optically-transparent current-conductive layer (for instance, ITO layer on glass substrate). In the case of the usage of Ti layer, which is also electrically conductive, after the laser processing, the nano-structured Ti layers are characterized by the periodic change of the nano-areas possessing both a non optically-transparent current-conductive layer Ti and an optically-transparent electrically non-conductive pure glass. Unfortunately, because of that periodic changes the nano-structured Ti layer is not appropriate for measured of the PAE $W_\theta$, because the unbroken conductive layer with periodic grooves should be used as recently it was realized for nano-structured ITO layer with periodic grooves [21]. On this basis of above stated, in this manuscript modified nano-structured Ti layer additionally coated with ITO layer was used and studied. In this case the unbroken nano-periodic conductive ITO layers, by reiterating the nano-profile of the processed Ti layer, are observed. A new modified aligning layer was obtained as a result. As the main parameters of the aligning layers, the AAE $W_\varphi$ and the PAE $W_\theta$ of pure ITO layer, already deposited on nano-structured Ti layer, and ITO layer additionally coated with polyimide PI2555 film were studied in this manuscript. Obviously, the different surfaces, treated by NLL with the use of various scanning speeds, and further modification of Ti layer by coating ITO and polymer layer, possess various quantities of AE, pretilt angle and contact angle of droplet of isotropic liquid and nematics.

**Experiment**

*2.1. Materials*

To study the alignment properties of the modified nano-structured Ti layers, two nematic liquid crystals, namely E7 and MLC-6609, obtained by Licrystal, Merck (Darmstadt, Germany) was chosen.

The optical and dielectrical anisotropy of the nematic E7 at T = 20 $^{o}$C, $\lambda$ = 589.3 nm, and $f$ = 1 kHz are $\Delta n$ = 0.2255 ($n_e$ = 1.7472, $n_o$ = 1.5217) and $\Delta \varepsilon$ = +13.8, respectively. Splay, twist and bend elastic constants of nematic E7 are $K_{11}$ = 11.7 pN, $K_{22}$ = 6.8 pN, $K_{33}$ = 17.8 pN, respectively [43-45]. The parallel and perpendicular components of electric permittivity of the E7 at T = 20 $^{o}$C is 19.5 and 5.2, respectively [46]. The temperature of the nematic-isotropic transition $T_{Iso}$ = 58 $^{o}$C [43].

To measure of the optical phase retardation of the VA LC cells the nematic LC MLC-6609 was used. The temperature of the nematic-isotropic transition $T_{Iso}$ = 91.5 $^{o}$C and rotational viscosity $\gamma_{rot}$ = 162 mPa·s [43,47,48]. The optical and dielectrical anisotropy of the nematic MLC-6609 at T = 20 $^{o}$C, $\lambda$ = 589.3 nm, and $f$ = 1 kHz are $\Delta n$ = 0.0777 ($n_e$ = 1.5514, $n_o$ = 1.4737) and $\Delta \varepsilon$ = -3.7 ($\varepsilon_\perp$ =7.1 and $\varepsilon_\parallel$ = 3.4), respectively.

Splay, twist and bend elastic constants of nematic MLC-6609 are $K_{11}$ = 17.2 pN, $K_{22}$ = 7.51 pN, $K_{33}$ = 17.9 pN, respectively [48].

To obtain the planar alignment in the azimuthal plane of the both nematics, n-methyl-2-pyrrolidone solution of the polyimide PI2555 (HD MicroSystems, USA) in proportion 10:1 was used. The refractive index of the polyimide of PI2555 is $n$ = 1.7 [49].

To study of the contact angle at the aligning layers the analytical reagent grade glycerol ($C_3H_8O_3$) with density $\rho$ = 1.256 g/cm$^3$ at T = 20 $^o$C and refraction index $n_D^{20}$ = 1.4728 was used.

## 2.2. The creation of modified nano-structured Ti layers

The creation of Ti layers was as follows. At the beginning 300 nm thick Ti layer was deposited on a glass substrate, as were recently described [20]. To create the 5×5 mm$^2$ area of nano-structured Ti layers we used the experimental scheme of the NLL [50]. Ti layers were treated by means of the NLL with various scanning speed $v$ at the constant LPF $J$ = 0.55 J/cm$^2$, resulting in various periodic nano-structured Ti layers, possessing the certain period Λ and depth A.

The next stage was a modification of the nano-structured Ti layers by coating of ~ 50 nm thick ITO layer on top of them. Such modified films were investigated as the first type of aligning layers (FTAL), schematically shown in Figure 2.

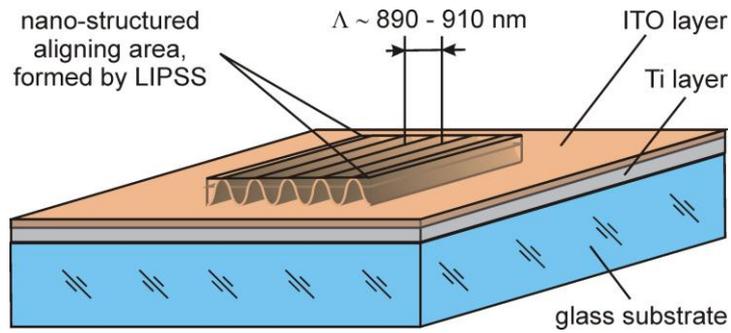

Figure 2. Schematic image of the nano-structured Ti layer treated by means of NLL, with area 5×5 mm² and additionally coated with ITO layer (FTAL).

## 2.3. Polymer film preparation

About 80 - 90 nm thick polyimide films at the surface of nano-structured Ti layers have been formed by means of dipping technique with the use of equipment for Langmuir-Blodgett film preparation R&K (Wiesbaden, Germany). For this purpose, substrates with the FTAL were dipped into prepared 10% n-methyl-2-pyrrolidone of the polyimide PI2555 solution followed by vertically lifting along the direction of the nano-grooves at the constant speed 2 mm/min. To evaporate the solvent, the polyimide PI2555 film was heated at $80^{\circ}$ C for 15 min and then the process of polymerisation was carried out at $180^{\circ}$ C for 30 min. Such modified films were used as the second type of aligning layers (STAL).

## 2.4. Methods

### 2.4.1. The determination of AAE $W_\varphi$

To determine AAE $W_\varphi$ the well-known method of combined twist LC cell [35,36], allowing to measure the twist angle $\varphi$, was used. For this aim combined twist LC cells, consisting of tested substrates, coated with modified films of both types, were made as recently described in [20]. The reference substrate was coated with the polyimide

PI2555 processed with the rubbing technique [19,20]. An experimental scheme to measure of twist angle φ of the combined twist LC cells is detailed elsewhere [20,35,36].

The measurement of twist angle $\varphi$ of the combined LC cell allows us to calculate the value AAE $W_\varphi$ of the aligning layer of tested substrate. The twist angle $\varphi$ is related to the AAE $W_\varphi$ as follows [36-38]:

$$W_\varphi = K_{22} \cdot \frac{2 \cdot \sin(\varphi)}{d \cdot \sin 2(\varphi_0 - \varphi)}, \qquad (1)$$

where $d$ is the thickness of the LC cell, $\varphi_0 \sim 36^o$ is the angle between the easy axes of the reference and tested substrates; $\varphi$ is the measured twist angle.

## 2.4.2. The measurement of LC pretilt angle $\theta_p$

Before the determination of PAE $W_\theta$ the LC pretilt angle $\theta_p$ measurements were carried out. The well known crystal rotation technique was used for this purpose and which is detailed elsewhere [51-54]. This method is to measure the optical transmission of LC cell as a function of the incident light angle. Here, LC cell was set between a polarizer and an analyzer, whose transmission axes are perpendicular to each other and make an angle of 45 degrees with the LC alignment direction. Therefore, the LC cell rotates around the perpendicular to the alignment direction axis. Consequently, the transmitted beam is measured for different angles of LC cell rotation. In this manuscript, the measurements of LC pretilt angle θp were carried out for LC cells, assembled with two the same substrates having either FTAL or STAL and filled by the nematic E7 ($\Delta\varepsilon > 0$) or MLC-6609 ($\Delta\varepsilon < 0$).

*2.4.3. The determination of PAE $W_\theta$*

To determine PAE $W_\theta$ that characterizes the director $\vec{n}$ deviations with respect to aligning layer normal, the electric field method [22-32], (well-checked for the various aligning layers processed by different techniques such as rubbing [21,29,30,32], photoalignment [32,55] and plasma beam [32]), has been applied.

Electric field method is well known as RV technique [29,30,33], which requires the measurement of phase retardation from voltage in contrast to classic method, knowing as Yokoyama-van Sprang technique [27] that requires the simultaneous measurement of the capacitance and optical phase. RV technique has a linear dependence, when the C is practically constant in some range of voltage ($V_{min}$, $V_{max}$). By knowing phase retardation $R_0$ at initial voltage (V = 0), it was obtained the formula, which allows determination of the PAE $W_\theta$ from a simple linear fit, as follows [29,30]:

$$\frac{R(V-V^*)}{R_0} = \tilde{J}_0 - \frac{2 \cdot K_1}{d \cdot W_\theta} \cdot (1 + \frac{K_3 - K_1}{K_1} \cdot \sin^2\theta_p) \cdot (V - V^*) \qquad (2)$$

where $V^* = \alpha \cdot \frac{\varepsilon_\parallel - \varepsilon_\perp}{\varepsilon_\parallel} \cdot V_{th}$, $V_{th} = \pi \cdot \sqrt{\frac{K_1}{\varepsilon_0 \cdot (\varepsilon_\parallel - \varepsilon_\perp)}}$, $\alpha$ is a coefficient in range ($2/\pi$, 1), $\tilde{J}_0$ is a constant, $K_1$ and $K_3$ is the splay and bend elastic constant of nematic LC, d is the thickness of LC cell, $\theta_p$ is the pretilt angle, $\varepsilon_\parallel$ and $\varepsilon_\perp$ represent parallel and perpendicular components of electric permittivity of LC.

The phase retardation was measured by means of simple Senarmont compensator [55] consisting of a polarizer P, sample – LC cell, $\lambda/4$ – quarter-wave plate (QWP) and an analyzer – A. In this scheme two Glan-Thompson prisms were used as a polarizer P and an analizer A. Each optical element of Senarmont compensator was characterized by a definite azimuth angle α$_i$, as shown in Figure 3.

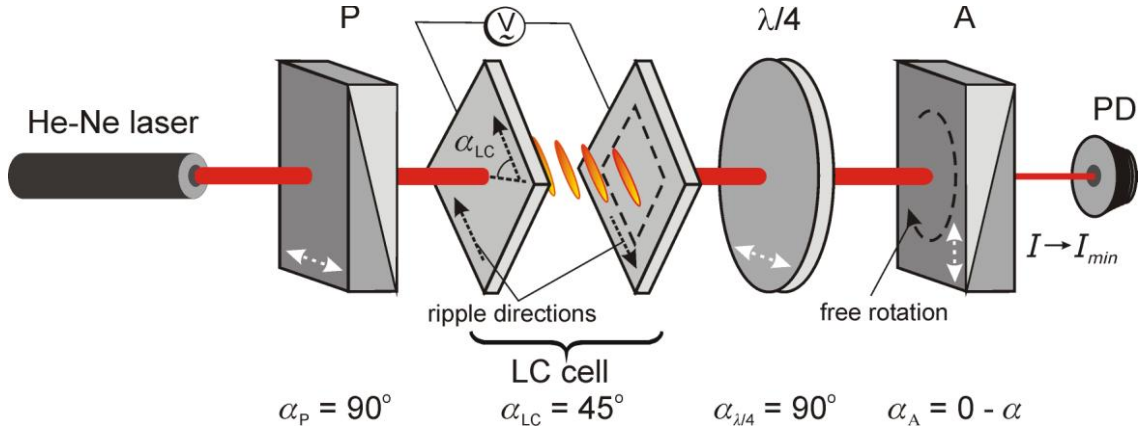

Figure 3. Optical scheme of Senarmont compensator, consisting of: He-Ne laser LGN-111 (λ = 632 nm and power P = 15 mW); P and A – polarizer and analyzer (Glan-Thompson prizms); Sample – symmetrical LC cell, consisting of a pair of substrates, possessing nano-structured Ti layers by means of NLL, additionally coated with ITO (FTAL), or FTAL additionally coated with polyimide PI2555 (STAL); λ/4 – quarter-wave plate (QWP); V - low frequency signal generator GZ-109; PD – silicon photodetector FD-18K, connected to the oscilloscope Hewlett Packard 54602B.

In this technique a sample was created as a symmetrical LC cell, where the orientation of nano-grooves (ripples) on both substrates was the same (Figure 2). LC cells, assembled from both FTAL and STAL, were studied by means of the Senarmont compensator. For this aim ripple directions on both substrates makes an angle α$_{LC}$ = 45° with the plane of polarizer P (white dashed arrows), as shown in Figure 3. The transmission axis of the polarizer P coincides with the optical axis of the quarter-wave (λ/4) plate (α$_P$ = α$_{λ/4}$ = 90°), as shown in Figure 3 by white dashed horizontal arrows. The transmission axis of the analyzer A is tunable by free rotating, within an angle range 0 – α.

The incident linearly polarized beam, by passing of the LC cell, becomes the elliptically polarized. This elliptical polarized beam converts back to the linear by means of the QWP placed after the LC, however with different polarization angle with respect to the input polarization before LC, depending on the phase retardation. By means of analizer A, we measured the azimuth of this linearly polarized beam that allows the determination of the optical phase retardation of LC cell [29,30]. For this purpose the analyser was rotated to find the angle α corresponding minimum of intensity at the photodetector PD.

*2.4.4. Determination of the contact angle β*

To measure the contact angle *β* of isotropic liquid (glycerol) or two nematic LCs (E7 and MLC-6609), the simple method has been used [40]. It is based on the measurement of linear dimensions of the droplet placed to the aligning layer by means of the horizontal microscope, as shown in Figure 4a.

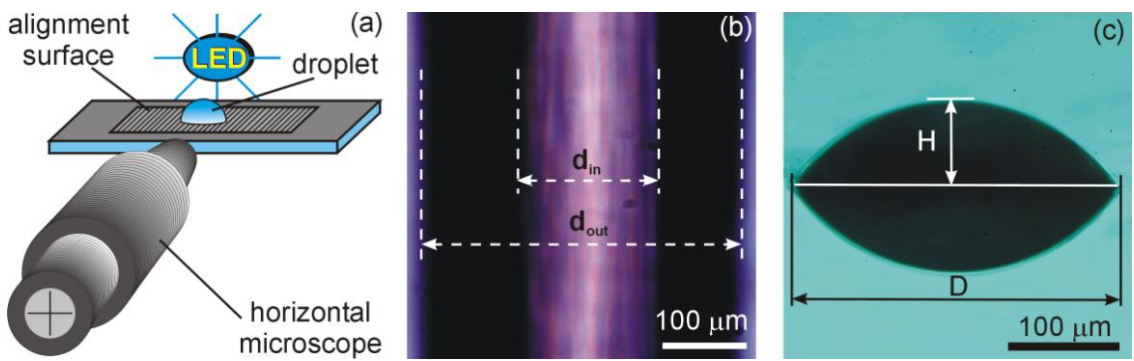

Figure 4. (a) The scheme to measure of the contact angle β, consisting of horizontal microscope with an ocular micrometer in an eyepiece, the studying aligning layer and deposited droplet (E7, MLC-6609 or glycerol). (b) The photography of capillary with geometrical dimensions, used for the deposition of droplets. The capillaries used possessed the outer diameter $d_{out}$ = 350 μm and inner opening with $d_{in}$ = 150 μm. (c). The photography of the glycerol droplet with certain diameter D and height H. The droplet was deposited on the pure nano-structured Ti layer (treated with scanning speed $\upsilon = 2000$ mm/s) and observed through an eyepiece of horizontal microscope.

Here, small size droplets of nematic LCs were deposited on various treated Ti surfaces by means of the tube capillary with outer and inner diameter 350 μm and 150 μm, respectively (Figure 4b). To ensure the accuracy of the measurement of the droplet dimensions, a light emitting diode (LED) lighting background was placed behind the droplet (Figure 4a), thereby increasing the contrast of the deposited droplet, as can be seen in Figure 4c. The measured values of diameter D and height H of droplet were used for the calculation of the contact angle $\beta$ according to the formula [57]:

$$\cos\beta = \frac{(D/2)^2 - H^2}{(D/2)^2 + H^2} \tag{3}$$

**Results and discussions**

*3.1. Azimuthal anchoring energy $W_\varphi$ of aligning surfaces*

*3.1.1. AFM studies of the modified aligning layers*

In this section the AAE $W_\varphi$ for two types of aligning layers will be studied in detail. However, in some cases, the FTAL and STAL under studied will be also compared with pure nano-structured Ti layer. At first, the estimation of the AE of various aligning surfaces by means of the Berreman theory [58,59] was carried out. The depth A and period Λ of nano-grooves learned from AFM studies should be used to this end. As an example, the typical AFM images are shown for the pure nano-structured Ti surface (Figure 5a), FTAL (Figure 5b) and STAL (Figure 5c). Here, the treatment Ti surface was carried out with scanning speed $v$ = 1500 mm/s and the laser pulse fluence (LPF) $J$ = 0.55 J/cm$^2$ and by other processes, namely the deposition of ITO and polyimide PI2555. The cross-section of pure nano-structured Ti surface, the FTAL and STAL is shown in Figure 5d, Figure 5e and Figure 5f, respectively.

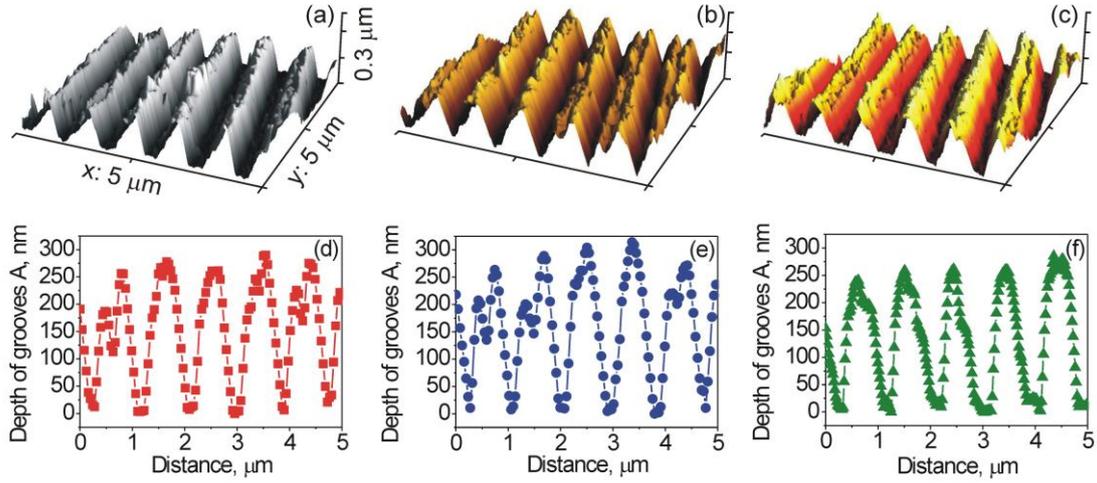

Figure 5. AFM images of the (a) pure nano-sturctured Ti layer, (b) pure nano-structuted Ti layer coated with the ITO film – FTAL and (c) pure nano-structured Ti layer covered the ITO film (FTAL) coated with the polyimide PI2555 film - STAL. The cross section of nano-grooves of the layer: (d) the pure nano-structured Ti, possessing average values of the period $\Lambda$ = 910 nm and depth A = 270 nm (solid red squares), (e) the FTAL, having average values of the period $\Lambda$ = 890 nm and depth A = 291 nm (solid blue circles) and (f) the STAL, possessing average values of the period $\Lambda$ = 890 nm and depth A = 256 nm (solid green triangles). Parameters of the NLL are: the scanning speed $v$ = 1500 mm/s and the laser pulse fluence $J$ = 0.55 J/cm$^2$.

The dependence of the nano-grooves depth A on the scanning speed $v$ at constant LPF $J$ = 0.55 J/cm$^2$ is shown in Figure 6. For the pure nano-structured Ti layer this dependence is a non-monotonic function (curve 1, solid red squares) as can be seen from Figure 6. However, it is Important to note that monotonous growing of the depth of grooves A has been observed when the process of the treatment of the Ti surfaces was carried out at a constant scanning speed $v$ for different value of LPF $J$, increasing its [19]. In our experiments the average value of period $\Lambda$ was about 910 nm, for the treated by NLL nano-structured Ti surfaces, which agree within 10 nm with values in [19,20].

In comparison with the pure structured Ti layers, the depth of grooves A of the FTAL is increased within range 20 - 40 nm, as can be seen from Figure 5e and Figure 6 (curve 2, opened blue circles). The average value of period of grooves $\Lambda$ of the FTAL was about 890 nm. The main reason of the decreasing of period of nano-grooves is the appearance of additional periodic structure after deposition of ITO layer on the structured Ti surface as can be seen from Figure 5 b and Figure 5e.

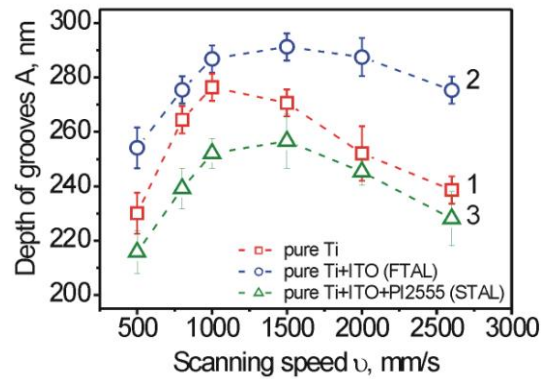

Figure 6. Dependence of the depth of nano-grooves A on the scanning speed $v$ for: 1) the pure nano-structured Ti layer (curve1, opened red squares); 2) the FTAL, consisting of the pure nano-structured Ti layer additionally coated with ITO (curve 2, opened blue squares); 3) the STAL, consisting of the FTAL additionally coated with polyimide PI2555 film (curve3, opened green treangles). During of the NLL technique of the Ti surface the LPF $J = 0.55$ J/cm$^2$. The dashed curve is a guide to the eye.

Conversely, the depth of grooves A for the STAL is decreased within range 30-50 nm in comparison with the FTAL as can be seen from Figure 5f and Figure 6 (curve

3, opened green triangles). It is therefore clear that the coating of the FTAL by polyimide PI2555 leads to the better filling of low-lying place of nano-grooves than their peaks. Due to this fact the depth of nano-grooves A has decreased but the period of grooves $\Lambda$ is the same as for the FTAL.

*3.1.2. The estimation of AE by Berreman theory*

By knowing values of the depth A and period $\Lambda$ of nano-grooves an estimation of AE $W_B$ by using the well-known Berreman's theory [58,59] can be carried out. The value of AE $W_B$ for the pure nano-structured Ti layer, the FTAL and the STAL was calculated as follows:

$$W_B = 2 \cdot \pi^3 \cdot K \cdot \frac{A^2}{\Lambda^3} \qquad (4)$$

where K is arithmetical mean of the all Frank constants ($K_{11}$, $K_{22}$ and $K_{33}$) of nematic LC.

The dependence of AE $W_B$ on scanning speed $v$ for the pure nano-structured Ti layer (curve 1, solid red squares), the FTAL (curve 2, solid blue circles) and the STAL (curve 3, solid green triangles) as an example for the nematic E7 are shown in Figure 7a.

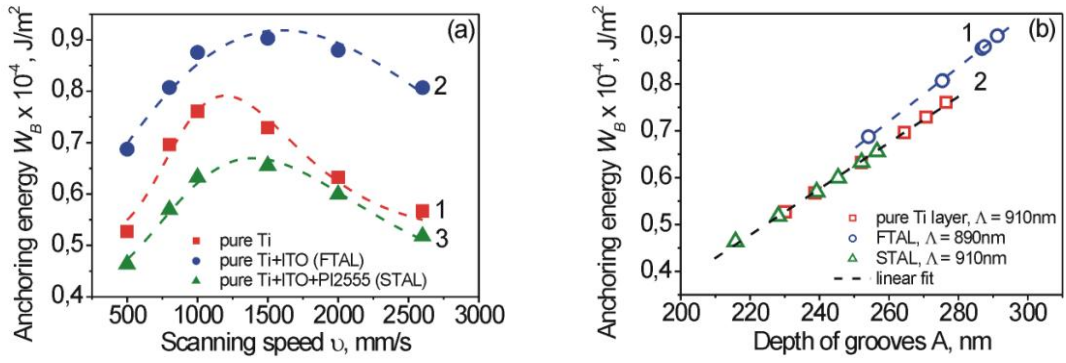

Figure 7. Dependence of AE $W_B$ on (a) the scanning speed $v$ and (b) depth of nano-grooves A for: 1) the pure nano-structured Ti layer (curve 1); 2) the FTAL, consisting of pure nano-structured Ti layer additionally coated with ITO (curve 2); 3) the STAL, consisting of the FTAL additionally coated with polyimide PI2555 film (curve 3). The value of LPF during of the NLL technique of the Ti surface was $J = 0.55$ J/cm$^2$. For calculation of AE $W_B$ the Frank constants of the nematic E7 were used. The dashed line is a guide to the eye.

It is important to note that the ITO layers coating the pure nano-strucrured Ti layers (so-called FTAL) lead to the certain increasing the depth of nano-grooves A and little change of period $\Lambda$ (Figure 5 and Figure 6).

The dependence of AE $W_B$ on depth of grooves A for the pure nano-structured Ti layers (opened red squares) and two modified layers, namely FTAL (opened blue circles) and STAL (opened green triangles) is shown in Figure 7b. Thus it can be concluded that some modification of pure nano-structured Ti layers such as coating by ITO and polyimide PI2555 layers can also change the depth of nano-grooves A and therefore the AE $W_B$. As can be seen from Figure 7a and Figure 7b the value of AE $W_B^{FLAT}$ for FTAL is larger than for the pure nano-structured Ti layer $W_B^{pure}$ and the STAL $W_B^{SLAT}$ (i.e. $W_B^{pure} < W_B^{FLAT}$ and $W_B^{SLAT} < W_B^{FLAT}$). It is seen, that for the FLAT the linear dependence of the AE $W_B(A)$ differ from that for pure nano-structured Ti layer and STAL. The reason of this is decreasing of the period of nano-grooves $\Lambda$ owing to the process of coating ITO layer to the nano-structured pure Ti layer. These data are in a good qualitatively agreement with the recently described calculations of AE $W_B$ [19,20].

It is seen that the STAL possessing the smallest values of depth of nano-grooves A, and thus of AE $W_B^{STAL}$ when compared with similar values for the pure nano-structured Ti layer and FTAL. As suggested earlier, the reason of the small values AE $W_B^{STAL}$ obtained for the STAL is a probably the unequal filling by polyimide PI2555 the low-lying places and peaks of nano-grooves. However, it should be noted that in [19,20] we recently found the difference between the calculated AE $W_B$ and experimentally obtained, using the method of combined twist LC cell [36,37], AAE $W_\varphi$,. One of the main reasons for this difference there is the fact that Berreman's theory has been not fully incorporating the physical and chemical interactions between LC molecules and surface [19,20]. As can be seen from Figure 7b (line 1, opened blue circles), the usage of any additional coating, which may increase the depth of nano-grooves A or/ and decrease of the period of nano-grooves Λ, will increase AE $W_B$.

*3.1.3. Determination of AAE $W_\varphi$ through the measurement of the twist angle $\varphi$*

Recently, it was shown that the AAE $W_\varphi$ of the treated by NLL surfaces could depend on, at the least, the two main parameters (e.g. LPS $J$ and scanning speed $v$), which use at the time of the creation of nano-structured layers [19,20]. In this manuscript the AAE $W_\varphi$ as a function of the scanning speed $v$ was also computed by twist angle $\varphi$ measurements, by using well-known technique of combined twist LC cell [51-54]. The measured twist angle $\varphi$ was thereafter used for the calculation of AAE $W_\varphi$ by equation (1).

The dependencies of twist angle $\varphi$ of LC cells filled by the nematic E7 ($\Delta\varepsilon > 0$) and the calculated AAE $W_\varphi$ on the scanning speed $\upsilon$ for various aligning layers are shown in Figure 8a and Figure 8b, respectively. The measurements of the twist angles $\varphi$ (Figure 8a) were carried out 30 min after the filling of LC cells by the nematic E7. It should be noted that for LC cells, consisting of pure nano-structured Ti layers (Figure 8, curves 1), the measured values of twist angles $\varphi$ (Figure 8a, opened red squares) and calculated AAE $W_\varphi$ (Figure 8b, solid red squares) are about the same as previously [19,20].

It is also important to note that for LC cells having the FTAL (Figure 8, curves 2) the transition from planar to homeotropic alignment (P-H transition) of nematic E7 over 2 hours was observed. In the case of the LC cells having pure nano-structured Ti layers (Figure 8, curves 1) or the STAL (Figure 8, curve 3) no P-H transitions were observed. The obvious conclusion is that the FTAL, consisting of the ITO layers, possess less surface tension $\gamma_S$ than tension of LC $\gamma_{LC}$ (*i.e.* $\gamma_S < \gamma_{LC}$) and as assumed by Creagh-Kmetz's rule [38,39], the P-H transition occurs.

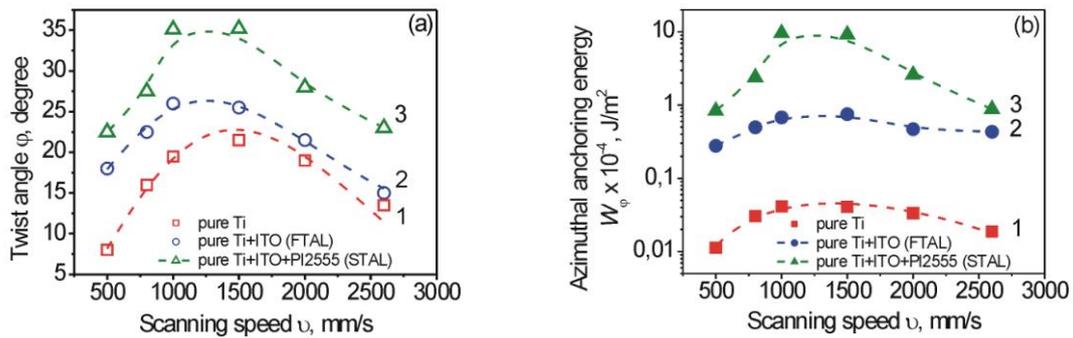

Figure 8. Dependencies of (a) twist angle $\varphi$ and (b) calculated AAE $W_\varphi$ of the LC cells, filled by nematic E7 ($\Delta\varepsilon > 0$), on scanning speed $\upsilon$ for the pure nano-structured Ti layer (curves 1), the FTAL, consisting of the pure nano-structured Ti layer additionally coated with ITO (curves 2) and the STAL, consisting of the FTAL additionally coated

with polyimide PI2555 (curves 3). Data from all curves (opened symbols) were measured in 30 min after filling of LC cells. The dashed line is a guide to the eye.

As can be seen from Figure 9a, for any of the aligning layers almost the linear dependence of calculated AAE $W_\varphi$ on the depth of the nano-grooves A measured by AFM is observed. It is clearly seen that the pure nano-structured Ti layers (Figure 9a, line 1), coated by both the ITO (Figure 9a, line 2) and polyimide PI2555 (Figure9a, line 3) leads to the increasing of AAE $W_\varphi$. In addition, for the STAL a strong AAE $W_\varphi$ can be observed even at their smaller depth of grooves A (Figure9a, line 3).

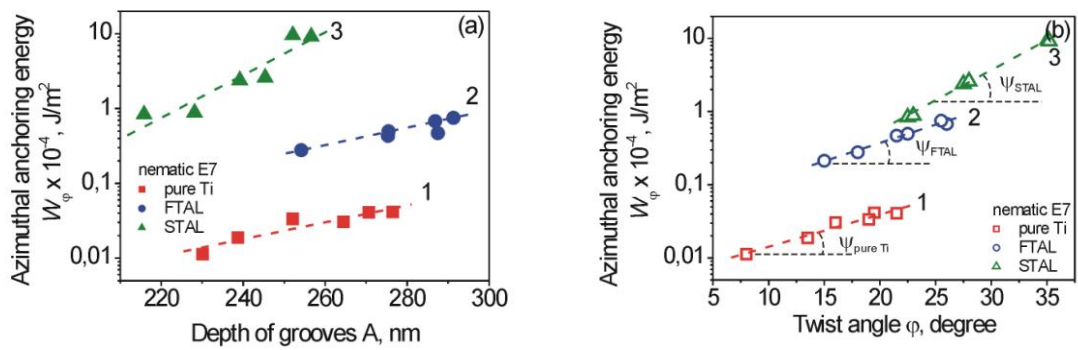

Figure 9. Dependence of the AAE $W_\varphi$ on: (a) the depth of nano-grooves A (solid symbols) and (b) the twist angle $\varphi$ (opened symbols) for various aligning layers: 1 – the pure nano-structured Ti layer, 2 – the FTAL and 3 – the STAL. The tilt angle $\psi$ of the linear fit 1, 2 and 3 is about 16, 18 and 29 degrees, respectively. LC cells were filled by the nematic E7 ($\Delta\varepsilon > 0$).

As can be seen from Figure 9a, the increasing of AAE $W_\varphi$ is strongly related to the rise in nano-grooves depth A. These dependencies are qualitatively coincident with those calculated previously [19,20] by means of the Berreman's theory.

However, as mentioned above, the value of the AAE $W_\varphi$ can also depend on the changing of physical and chemical properties of the surface. Here, the modification of aligning layers was realized through the usage of various materials, namely ITO and PI2555. Recently, it was concluded that the increasing of AAE $W_\varphi$ could be gained by the usage of polymer having stronger anchoring [19,20]. In this manuscript, in contrast to recently used polyimide ODAPI, having AAE $W_\varphi \sim 1\times10^{-4}$ J/m$^2$ [19,20], polyimide PI2555 with essential strong AAE $W_\varphi \sim 9\times10^{-4}$ J/m$^2$ was applied. It is obvious that reason of this fact can be explained as follows. For example, non-rubbed polyimide PI2555 layer also possesses a strong AAE $W_\varphi$ in comparison with the same untreated ODAPI layer, since under identical conditions of rubbing technique value of $W_\varphi$ for PI2555 is about $4\times10^{-4}$ J/m$^2$ [60,61], while for ODAPI $W_\varphi \sim 1\times10^{-4}$ J/m$^2$ [62].

In Figure 9b the dependence of AAE $W_\varphi$ on twist angle $\varphi$ is shown. This dependence is a linear function with a certain slope $\psi$ (Fig. 9b). Interestingly, for various aligning layers the slopes $\psi$ are different and form about 16, 18 and 30 degrees for the pure nano-structured Ti layer, FTAL and STAL, respectively. As shown above, the presence of the additional layers obtained by coating on the surface of nano-structured Ti layer different materials (*e. g.* ITO or PI2555) leads to a change in both the geometrical parameters of nano-grooves (depth A and period Λ) and physical and chemical characteristics of the aligning layers. Such changes increase the twist angle $\varphi$, and lead to growing of the AAE $W_\varphi$, (Figure 9b). For example, the coating of ITO onto the pure nano-structured Ti layer (the FTAL, line 2 and solid blue circles) increases the twist angle to $\varphi \approx 25$ degree and therefore the AAE to $W_\varphi \approx 0.9\times10^{-4}$ J/m$^2$. While further coating of PI2555 onto the FTAL (STAL, line 3 and solid green triangles) leads to even greater values of these characteristics, namely $\varphi \approx 35.5$ degree and $W_\varphi \approx 9\times10^{-4}$ J/m$^2$.

Similar dependencies $\varphi(v)$ and $W_\varphi(v)$ for LC cells filled by the nematic MLC-6609 ($\Delta\varepsilon < 0$) are shown in Figure 10. Here, in comparison to the nematic E7 for the pure nano-structured Ti layers [19,20], homeotropic alignment was observed after the filling LC cells by the nematic MLC-6609. The twist angles $\varphi$ and the calculated AAEs $W_\varphi$ as functions of the scanning speed $v$ for the FTAL (curve 1) and the STAL (curve 2) are shown in Figure 10a and Figure 10b, respectively.

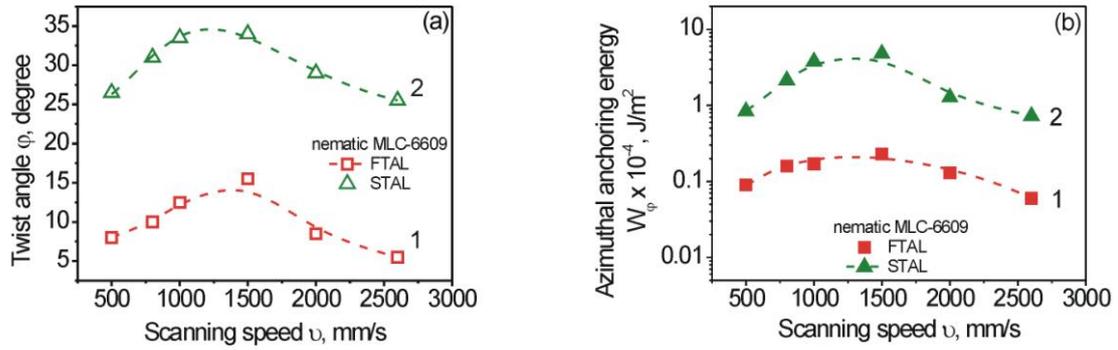

Figure 10. Dependence of (a) the twist angle $\varphi$ and (b) the calculated AAE $W_\varphi$ of the LC cells, filled by the nematic MLC-6609 ($\Delta\varepsilon > 0$), on the scanning speed $v$ for the FTAL (curves 1), consisting of the pure nano-structured Ti layer additionally coated with ITO, and the STAL (curves 2), consisting of the FTAL additionally coated with the polyimide PI2555. The dashed line is a guide to the eye.

It should also be mentioned that similarly to LC cells filled by the nematic E7 ($\Delta\varepsilon > 0$), for combined twist LC cells, consisting of the FTAL (curve 1) and filled by the nematic MLC-6609, the P-H transition occurs within 2 hours. Therefore, by using FTAL (curve 1) the planar alignment of the MLC-6609 will be always unstable. Because of this finding, the LC cells are unable to rotate the linearly polarized light similarly to combined twist LC cells described elsewhere [36,37,62]. Consequently, it should be noted that in the case of FTAL, the measurements of the twist angles $\varphi$ were carried out immediately after filling of LC cells.

As can be seen from Figure 8 and Figure 10, the usage of additional layers (*e. g.* ITO and PI2555) leads to the increasing both the twist angle $\varphi$ and the AAE $W_\varphi$. For instance, it can be seen from Figure 8a that for the combined twist LC cells, consisting of the pure nano-structured Ti layers (curve 1, opened red squares) and filled by the nematic E7, the weak rotation of the linearly polarized light (small values of twist angle $\varphi$) occurs. Furthermore, for similar LC cells filled by the nematic MLC-6609 ($\Delta\varepsilon < 0$) no rotation of linearly polarized light is observed due to the VA of LC molecules. Unlike this layer, for FTAL (curves 1) and STAL (curves 2), coated with both ITO and polyimide PI2555, the increasing of twist angle $\varphi$ and AAE $W_\varphi$ is observed. Finally, it is important to note that additional PI2555 layer leads to stable planar alignment of both nematics throughout the storage time. As was already mentioned above, the coating with ITO (FTAL) is the reason for unstable planar alignment of both E7 ($\Delta\varepsilon > 0$) and MLC-6609 ($\Delta\varepsilon < 0$) and their further P-H transition. This may be explained, for example, by the interaction between the surface and LC or in other words, by the different ratio between surface tension of an aligning layer $\gamma_S$ and tension of LC $\gamma_{LC}$. In addition to this, let us recall that both parameters, namely AAE $W_\varphi$ and PAE $W_\theta$ and their ratio are also important characteristics of aligning surfaces.

### 3.2. Polar anchoring energy $W_\theta$ of aligning surfaces

In this section the influence of the scanning speed during treatment of Ti surface and their modification by different layers (*e.g.* ITO and polyimide PI2555) on the pretilt angle $\theta_p$ and PAE $W_\theta$ of symmetrical LC cells would be investigated. The LC cell was consisted of a pair of substrates coated with FTAL or STAL and filled by nematics E7 ($\Delta\varepsilon > 0$) or MLC-6609 ($\Delta\varepsilon < 0$).

It is well known that for the calculation of PAE $W_\theta$, different methods [27-31,33] can be used. The methods proposed involve studying of LC cells under the action of an alternating electric field. As mentioned above in the Introduction section, because of this fact, the LC cells consisting of pair pure nano-structured Ti layers will not be examined, because these LC cells not conduct of the current over the whole treated surface area.

*3.2.1. The measurement of pretilt angle $θ_p$*

In Figure 11 the dependence of pretilt angle $θ_p$ both the FTAL and the STAL on the scanning speed $υ$ are shown.

Let us recall that the coating of ITO onto pure nano-structured Ti layer leads to the increase of the depth of nano-grooves A (Figure 7a). For symmetrical LC cells, consisting of two substrates with FTAL layer and filled by nematic E7 (or MLC-6609), there are non-monotonic dependencies of the pretilt angle $θ_p$ on scanning speed $υ$. As can be seen from Figure 11a the pretilt angle $θ_p$ of LC cells, filled by the nematic E7 (curve 1-0, solid blue squares), is about 1.5 times lower than for the nematic MLC-6609 (curve 2-0, solid red circles). Such a difference can mainly be explained by various values of AAE $W_φ$, PAE $W_θ$ and their ratio due to the different interaction between the aligning surface and LC. It can be assumed that the surface possessing strong AAE $W_φ$ will have less pretilt angle $θ_p$ of LC and vice versa. This assumption can be experimentally confirmed by comparing pretilt angle $θ_p$ (curve 1-0, Figure 11a or curve 2-0, Figure 11a) with AAE $W_φ$ (curve 2, Figure 8b or curve 1, Figure 10b) for LC cells, filled by the nematic E7 and MLC-6609, respectively. It is easy seen that for each nematic LC these dependencies possess opposite to each other extrema within the scanning speed range 1000 – 1500 mm/s.

It was found that in the case of FTAL the pretilt angle $θ_p$ both the nematic E7 and the MLC-6609 is unstable and changes over time, as shown in the Figure 11b (dashed red arrow). The P-H transition of both nematics lasted a total of two hours. The inset of the Figure 11b depicts the dependence of the pretilt angle $\theta_p$ on the scanning speed $v$ for the homeotropically oriented nematic E7 (curve 1, opened blue squares) and MLC-6609 (curve 2, opened red circles). It is seen that the pretilt angles $\theta_p$ are within the range of 86 – 89 degrees.

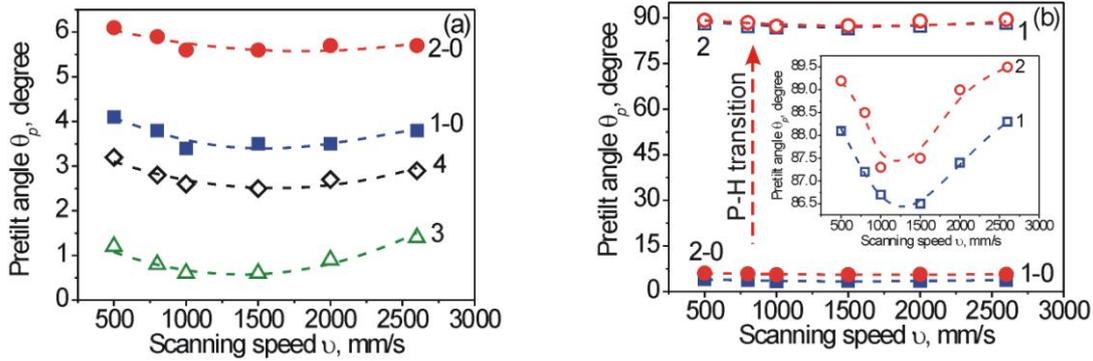

Figure 11. Dependence of the pretilt angle $\theta_p$ of the symmetrical LC cell on the scanning speed $v$ for: (a) the unstable planar alignment of the nematic E7 (curve 1-0, solid blue squares) and MLC-6609 (curve 2-0, solid red circles) for the FTAL. The stable planar alignment of the nematic E7 (curve 3, opened green triangles) and MLC-6609 (curve 4, solid black diamonds) for the STAL; (b) the P-H transition of the nematic E7 and MLC-6609 for the FTAL. The curve 1-0 (solid blue squares) and the curve 2-0 (solid red circles) - the pretilt of the planar alignment of the nematic E7 and MLC-6609, respectively. The curve 1 (opened blue squares) and the curve 2 (opened red circles) - the pretilt homeotropic alignment of the nematic E7 and MLC-6609, respectively. The inset depicts the $\theta_p(v)$ for the FTAL after the P-H transition in a large scale. The dashed line is a guide to the eye.

However, the usage of additional layer of polyimide PI2555, coating onto the FTAL, leads to the fact that the pretilt angles $\theta_p$ of the nematic E7 (Figure 11a, curve 3, opened green triangles) and the nematic MLC-6609 (Figure 11a, curve 4, opened black diamonds) do not change over time. Here, as in the case of the FTAL, for the both nematics the non-monotonic dependence of pretilt angle $\theta_p$ on scanning speed $v$ is observed. As can be seen from Figure 11a for STAL the pretilt angles $\theta_p$ of the nematic

E7 (or MLC-6609) for a certain scanning speed $v$ are smaller in comparing with the pretilt angles $\theta_p$ for the same speed in the case of FTAL. The reason of the small values of the pretilt angle $\theta_p$ is a strong AAE $W_\varphi$ and obviously weak PAE $W_\theta$ of the polyimide PI2555 layers applying for the planar alignment of LCs [63].

*3.2.1. The measurement of the optical phase retardation. The calculation of PAE $W_\theta$*

In order to calculate the PAE $W_\theta$, two different methods of the optical phase retardation measurements were used [29,30,33]. To measure of optical phase retardation of FTAL the optical scheme for VA of LC cell filled by nematic MLC-6609 with negative $\Delta\varepsilon < 0$ was used [33]. In case of STAL the simple Senarmont compensator [56] was used in order to measure of the optical phase retardation of the LC cell possessing the planar alignment as detailed previously [29,30,32].

The typical optical phase retardation function $R(V-V^*)$ of LC cells from applied voltage $(V-V^*)$, which was used to calculate PAE $W_\theta$ of aligning layers is shown in Figure 12. As can be seen from Figure 12, in the certain voltage range $(V_{min}, V_{max})$ the optical retardation as a function of applied voltage is a linearly dependent ($y = A + B \cdot x$).

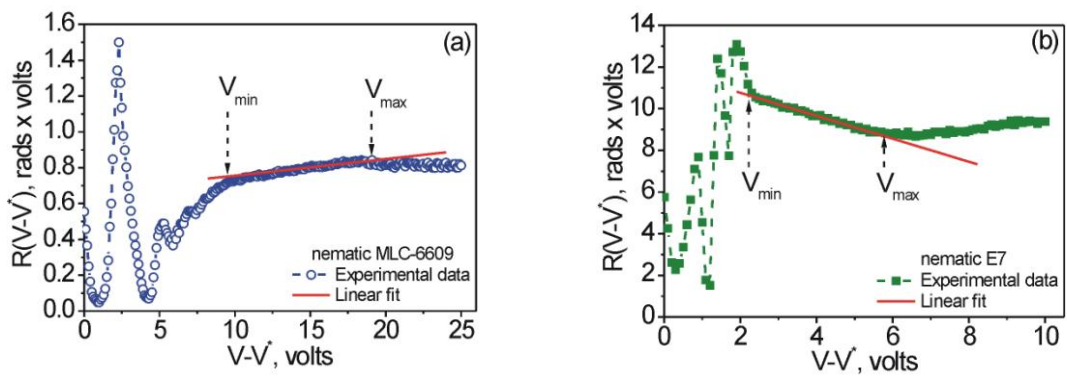

Figure 12. Dependence of the optical phase retardation function $R(V-V^*)$ on $(V-V^*)$ of the LC cell consisting with: (a) FTAL and filling by the nematic MLC 6609 ($\Delta\varepsilon < 0$) and (b) STAL and filling by the nematic E7 ($\Delta\varepsilon > 0$). For both types of the aligning layers, the scanning speed $\upsilon$ and LPF $J$ were 500 mm/s and 0.55 J/cm$^2$, respectively. The thickness of LC cell was 23.5 μm (a) and 25.7 μm (b). The values of the linear fit parameters are: (a) A = 11.8649, B = - 0.5512, $V_{min}$ = 9.7V and $V_{max}$ = 17.9V, (b) A = 0.6628, B = 0.0093, $V_{min}$ = 2.3V and $V_{max}$ = 5.7V. Estimated values of the PAE $W_\theta$ for (a) FTAL and (b) STAL are about $0.7\times10^{-4}$ and $0.2\times10^{-4}$ J/m$^2$, respectively.

As an example, two linear dependencies of R(V-V*) on (V-V*) for FTAL and STAL are shown Figure 12a and Figure 12b, respectively. Ti surfaces for FTAL and STAL were treated by NLL under the same conditions, namely the scanning speed $\upsilon$ and the LFT $J$ were 500mm/s and 0.55 J/cm$^2$, respectively. The calculated values of the PAE $W_\theta$ of FTAL and STAL are ~ $0.7\times10^{-4}$ J/m$^2$ and $0.2\times10^{-4}$ J/m$^2$, respectively.

By using the linear range of the optical phase retardation, the PAE $W_\theta$ were calculated for both FTAL and STAL, made on the basis of nano-structured Ti layers, which were obtained by NLL at various scanning speeds $\upsilon$. The dependencies of PAE $W_\theta$ on the various scanning speeds $\upsilon$ are shown in Figure 13a.

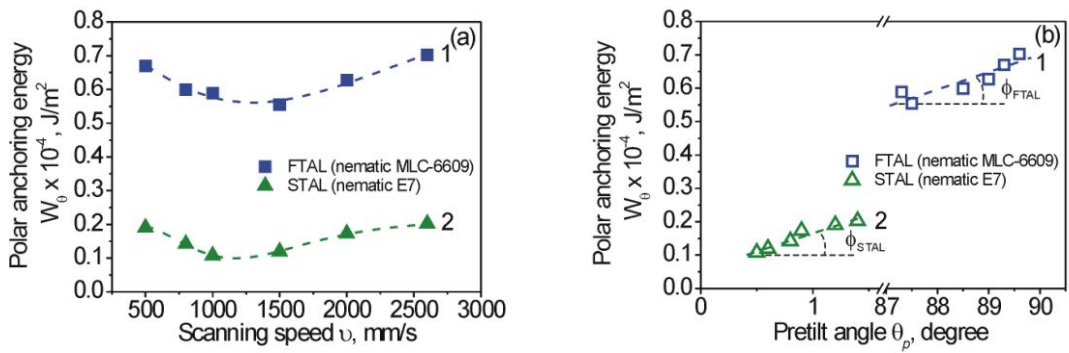

Figure 13. Dependencies of the calculated PAE $W_\theta$ on: (a) the scanning speed $\upsilon$ and (b) pretilt angle $\theta_p$. LC cell consists of two aligning layers: FTAL (curve 1, solid blue squares) or STAL (curve 2, solid green triangles). LC cells were filled by the nematic MLC-6609 (curve 1) and E7 (curve 2), respectively. The slopes $\phi$ for the FTAL (line 1,

opened blue squares) and the STAL (line 2, opened green triangles) are 86 and 78 degrees, respectively. The dashed line is a guide to the eye.

As was mentioned above, for FTAL the P-H transition both the nematic E7 ($\Delta\varepsilon > 0$) and MLC-6609 ($\Delta\varepsilon < 0$) is observed. Due to this fact, the measurement of optical phase retardation from applied field was possible only upon usage the VA of the MLC-6609, possessing $\Delta\varepsilon < 0$ (Figure 12a). The dependence of PAE $W_\theta$, on the scanning speed $v$, for this case is shown in Figure 13a (curve 1, solid blue squares).

In contrary, the usage of STAL leads to the planar alignment for both E7 ($\Delta\varepsilon > 0$) and MLC-6609 ($\Delta\varepsilon < 0$) nematics. The reason of this planar alignment is the availability of the additional polyimide PI 2555 layer that results in stronger AAE $W_\varphi$, compare to the one observed for FTAL. In this case of the measurement of optical phase retardation from applied alternating field was possible only for the nematic E7, having $\Delta\varepsilon > 0$ (Figure 12b). The dependence of PAE $W_\theta$, on scanning speed $v$ for planar alignment of the nematic E7 is shown in Figure 13a (curve 2, solid green triangles).

As can be seen from Figure 13a, there exist extremes (minimums) of the $W_\theta(\theta_p)$ within the scanning speed range 1000 – 1500 mm/s for both FLAT and SLAT. In addition, by taking into account the fact that both the depth of the nano-grooves A (Figure 6) and the PAE $W_\theta$ (Figure 13a) are depend on scanning speed $v$, it can be readily seen the correlation between the nano-grooves depth A and the PAE $W_\theta$. For instance, for the scanning speeds $v$ in which nano-structures possessing a smaller nano-grooves depth A (Figure 6, curves 2 and 3) PAE $W_\theta$ are stronger (Figure 13a, curves 1 and 2) and vice versa.

In Figure 13b for the both aligning layers the linear dependence of calculated PAE $W_\theta$ on pretilt angle $\theta_p$ is shown. As can be seen from Figure 13b for FTAL (line 1, opened blue squares) the slope of the line $\phi_{FTAL}$ ~ 86 degrees, while for STAL (line 2, opened green triangles) $\phi_{STAL}$ is about 78 degrees. The difference may be explained by the fact that STAL possesses a weak PAE $W_\theta$ (Figure 13a, curve 2) due to the polyimide PI2555 layer, characterized by strong AAE $W_\varphi$ [60,61]. Let us recall, a strong AAE there is the reason for decreased pretilt angle $\theta_p$ (Figure 11a, curve 3). As can be seen from Figure 13b the increasing of pretilt angle $\theta_p$ leads to PAE $W_\theta$ growth. The inclination angle $\phi$ of linear dependence $W_\theta$ on $\theta_p$ may be seen as a typical parameter, which characterizes a certain type of aligning layer.

In addition, under the certain experimental conditions some correlation between PAE $W_\theta$ and AAE $W_\varphi$ is observed. In contrary to [63], here for example, nano-structured surfaces possessing relatively strong PAE $W_\theta$ (Figure 13a, curve 1) have a weak AAE $W_\varphi$ (Figure 8b, curve 2), and vice versa. The same correlation between AAE $W_\varphi$ and PAE $W_\theta$ were recently found during their measurements and studies of easy orientation axis gliding in a magnetic field on photoaligning PVCN-F layer [64]. Here authors found that relatively weak AAE $W_\varphi = (10^{-7} – 10^{-5})$ J/m$^2$ leads to strong drift of easy orientational axis when PAE $W_\theta = 10^{-4}$ J/m$^2$ in a strong electric field.

As can be seen from Figure 13a, the coating of the PI2555 layer onto the FTAL leads to weaker PAE $W_\theta$ of the formed STAL (curve 2). Here, as already mentioned above, the main reason of stronger AAE $W_\varphi$ of STAL (Figure 8b, curve 3) is polyimide PI2555 layer, possessing strong AAE $W_\varphi$ in comparison with FTAL (Figure 8b, curve 2). This correlation between PAE $W_\theta$ and AAE $W_\varphi$ can also be understood from the schematic presentation of distribution of the director $\vec{n}$ in the bulk of LC, which is bordered on the nano-structured aligning surface (Figure 1).

The dependence of PAE $W_\theta$ on AAE $W_\varphi$ both FTAL (opened blue diamonds) and STAL (solid green spheres) is also shown in Figure 14, when LC cells were filled by the nematic MLC-6609 ($\Delta\varepsilon < 0$) and E7 ($\Delta\varepsilon > 0$), respectively.

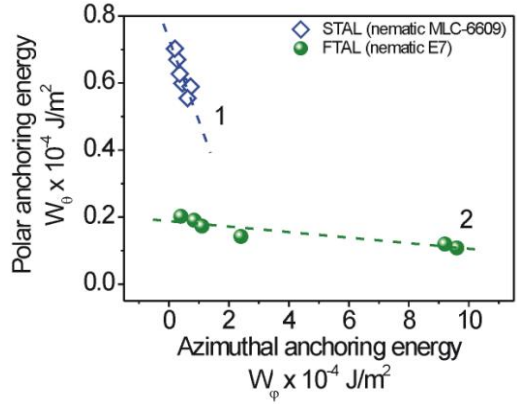

Figure 14. The correlation between PAE $W_\theta$ and AAE $W_\varphi$ for: FTAL (opened blue diamonds) and STAL (solid green spheres). LC cells were filled by the nematic MLC-6609 (line 1) and E7 (line 2).

It is seen that under certain conditions, namely when PAE $W_\theta$ of the aligning layer is increasing (or pretilt angle $\theta_p$ of the nematic MLC-6609 is decreasing) the decreasing of AAE $W_\varphi$ of this layer is observed. For instance, in the case of various FTAL (opened blue diamonds), which differ by scanning speed $\upsilon$ during the NLL processing, VA of the nematic MLC-6609 with small pretilt angle $\theta_p$ (close to 90 degrees) occurs. As can be also seen from Figure 14, for the STAL (solid green spheres), having the polyimide PI2555 layers, the strong AAE $W_\varphi$ is observed. This results in planar alignments of the nematic E7.

### 3.3. Contact angle β of nematic droplets on the aligning layers

It is assumed that one main reason for wettability of solid surfaces is both the interaction between interfacial tension of all materials which make contact and the geometrical properties of the surface (*i. e.* random roughness or periodical nano-grooves with period Λ and depth A). By taking into account of the empirical Creagh-Kmetz's rule [38,39], here, some correlation between the contact angle $β$ of nematic droplets deposited on FTAL and STAL, and calculated AAE $W_φ$ (or/and PAE $W_θ$) will be described.

*3.3.1. Contact angle of the droplets on rubbed polymer films*

First of all, in order to demonstrate the impact of both various polymers and different processing methods on contact angle $β$ of droplets (*e.g.* two nematics and glycerol), we have chosen the polyimide PI2555 and ODAPI, having different values of AAE $W_φ$ (PAE $W_θ$) [60-62].

At the beginning, the dependence of the droplets' contact angle $β$ for two nematics and glycerol on the number of rubbing $N_{rubb}$ for the polyimide PI2555 and ODAPI layers was studied. As can be seen from Figure 15, for droplets of the two nematics and glycerol the non-monotonic dependence $β(N_{rubb})$ for each polymer layers is observed. In this case all the measurements of geometrical dimension of droplets were carried out along direction of rubbing. In the case of ODAPI, a certain correlation between the contact angle $β$ (Figure 15a, curve 2) and calculated value of AAE $W_φ$ [62] for the number of rubbing $N_{rubb}$ is observed, similarly to the case of the photosensitive PVCN-F surface [40]. Consequently, for the ODAPI layers possessing a strong AAE $W_φ$ (when $N_{rubb}$ = 10 - 15), small contact angles $β$ were found. As can be seen from Figure 15a for the glycerol (curve 1, opened red squares) and two nematics, namely the E7 (curve 2, opened blue circles) and the MLC-6609 (curve 3, opened green triangles), the contact angle $β$ increasing for ODAPI layers having a weak $W_φ$.

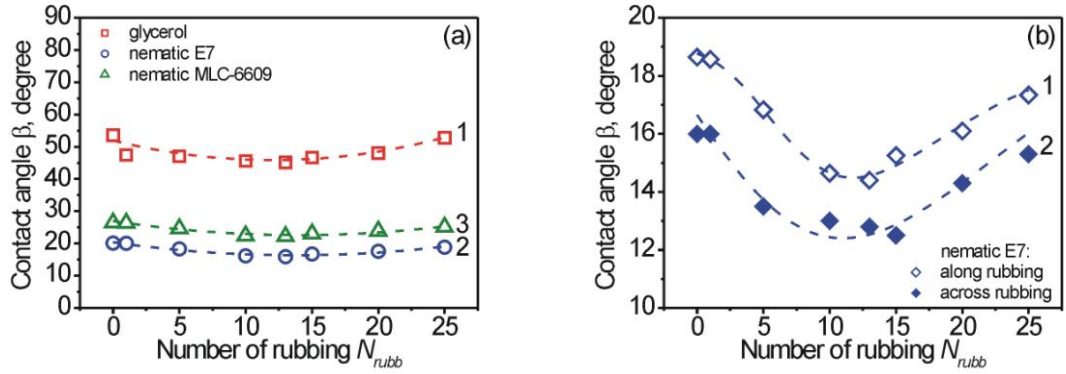

Figure 15. Dependence of the contact angle β on the number of rubbing $N_{rubb}$ for: (a) polyimide ODAPI layer for droplet of glycerol (curves 1, opened red squares), nematic E7 (curves 2, opened blue circles) and MLC-6609 (curves 3, opened green triangles), measured along direction of rubbing and; (b) polyimide PI2555 layer for droplet of nematic E7, measured along (curve 1, opened blue diamonds) and across (curve 2, solid blue diamonds) direction of rubbing. The dashed curves are guide to the eye.

In addition, it was also found that for the same treated surface the measured contact angles β (or geometrical dimensions of droplets D and H) along and across to the direction of rubbing of the polymer have different values. The dependence of the contact angle β on the number of rubbing $N_{rubb}$, for the PI2555 layer measured along (opened squares) and across (solid squares) to the direction of rubbing is shown in Figure 15b. It is seen that the droplet is expanding more in the direction of rubbing compare to the transverse the direction (or, in other words, the expanding is observed along the periodic grooves, obtained after the process of the rubbing). Here, it is important to note that contrary to both nematic droplets, for glycerol droplets the dimensions both the diameter D and the height H are not changed. The main reason is that the glycerol is an isotropic liquid.

*3.3.2. Contact angle of droplets on modified nano-structured surfaces*

It may be assumed that the aligning layer created by NLL and further coated by ITO and polyimide PI2555 will have similar properties as in case of the usage of rubbing technique. Recently, it was shown that the graphene surfaces, differently treated by changing the laser power during the NLL process, possessing various values of the contact angles $\beta$ of droplets deposited on those [42]. Here, we present the results of the droplets' contact angle measurements on the surfaces obtained by NLL with different scanning speeds $\upsilon$ and modified by both ITO and polymer PI2555 coating.

At the photograph (Figure 16) droplets of glycerol and both nematics, which were deposited on FTAL surface are shown. For this experimental conditions the FTAL possessed the certain value of the AAE $W_\varphi \sim 0.47 \times 10^{-4}$ J/m$^2$ and Ti layer was treated at the scanning speed $\upsilon = 2000$ mm/s. As can be seen from Figure 16 the values of contact angles $\beta$ for various droplets are different. For isotropic liquid (glycerol) the value of the contact angle $\beta$ is ~ 53 degree, while for the droplets of nematic E7 and MLC-6609 values of contact angles $\beta$ are ~ 14 and 27 degrees, respectively.

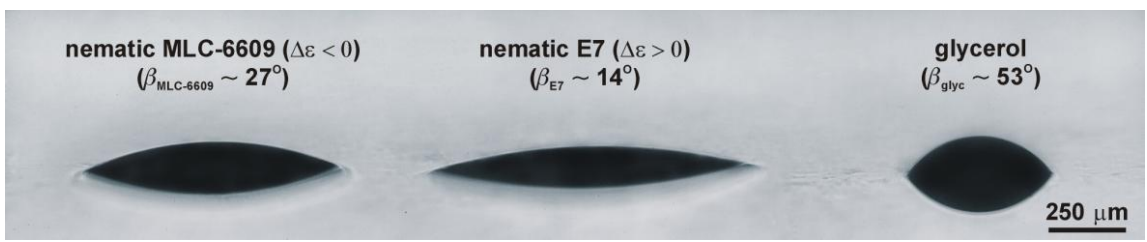

Figure 16. Photography of droplet of glycerol and two nematics, namely MLC-6609 and E7, placed at the aligning surface FTAL, obtained under NLL treatment of the Ti layer with the laser pulse fluence $J = 0.55$ J/cm$^2$ and the scanning speed $\upsilon = 2000$ mm/s with further modification by means of the ITO-coating. FTAL possesses by nano-grooves with average values of the period $\Lambda = 890$ nm and depth A = 291 nm. The value of the AAE $W_\varphi$ was about $0.47 \times 10^{-4}$ J/m$^2$.

The dependence of the contact angle $\beta$ for glycerol and two nematics droplets, deposited on both FTAL and STAL, on scanning speed $v$ is shown in Figure 17a and Figure 17b, respectively.

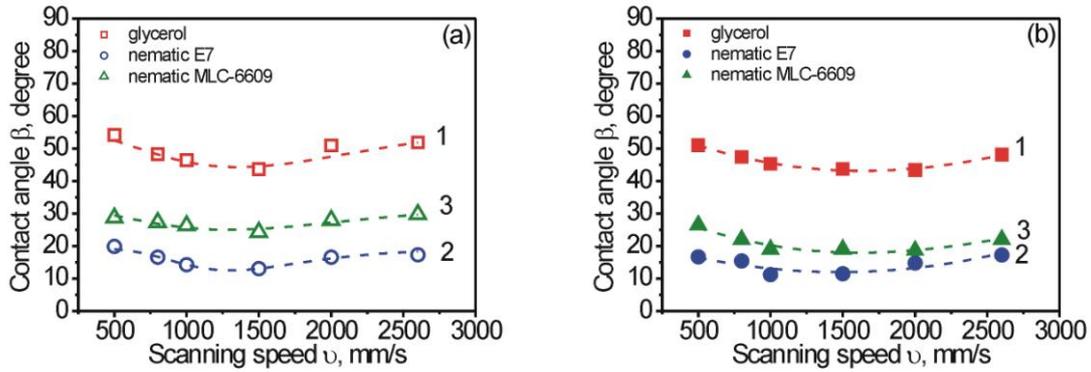

Figure 17. Dependence of the contact angle $\beta$ of droplet of the glycerol (curves 1), nematic E7 (curves 2) and nematic MLC-6609 (curves 3) deposited on (a) FTAL and (b) STAL, on scanning speed $v$ during process of NLL. Contact angles $\beta$ were measured along the rubbing direction. The dashed curve is a guide to the eye.

As can be seen from Figure 17, for all liquids, deposited on the FTAL, contact angles $\beta$ are more than in case of STAL under the same conditions of treatment of Ti surface by means of NLL. It can be assumed that FTAL has low wettability of the surface than STAL, because STAL consists of the polyimide PI2555 layer possessing strong AAE $W_\varphi$. As noted above, the aligning layer, having strong AAE $W_\varphi$, characterized by weak PAE $W_\theta$ and vice versa [64]. It is obvious that certain correlation between the contact angle $\beta$ and AAE $W_\varphi$ (and/ or PAE $W_\theta$) may also be seen. Dependence of the calculated AAE $W_\varphi$ (lines 1) and PAE $W_\theta$ (lines 2) on the contact angle $\beta$ for nematic droplet MLC-6609 deposited on FTAL and nematic droplet E7 deposited on STAL are shown in Figure 18a and Figure 18b, respectively.

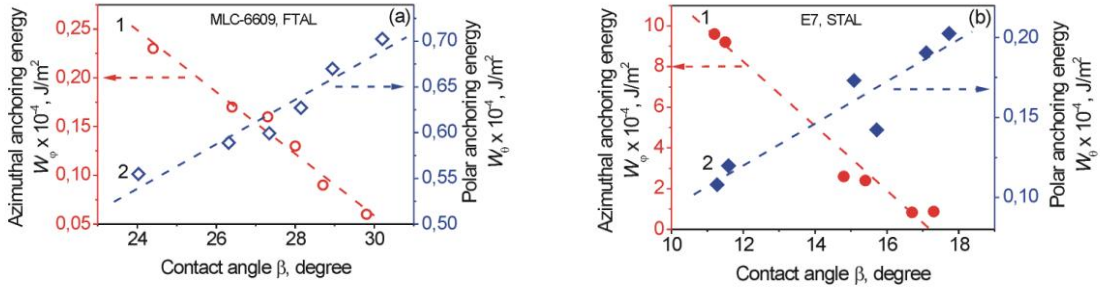

Figure 18. Dependence of the calculated values of the AAE $W_\varphi$ (lines 1) and the PAE $W_\theta$ (lines 2) on contact angle $\beta$ of the droplet of: (a) nematic MLC-6609 deposited on FTAL and (b) nematic E7 deposited on STAL. Contact angles $\beta$ were measured along the rubbing direction. The dashed line is a guide to the eye.

It is easy to see that the decreasing of AAE $W_\varphi$ (lines 1) resulting in the increasing of the contact angle $\beta$ for each nematic droplets when using both FTAL (Figure 18a) and STAL (Figure 18b). In addition, for FTAL, having a stronger PAE $W_\theta$ (Figure 18a, line 2 and opened blue diamonds), the contact angle $\beta$ is more than for STAL, possessing a weak PAE $W_\theta$ (Figure 18b, line 2 and solid blue diamonds).

**Conclusions**

In this manuscript, for the first time, we studied in detail the aligning surfaces, obtained by the NLL method, which was recently proposed [19,20] with further additional ITO and polyimide PI2555 films coating. For various aligning layers such as FTAL and STAL the dependence of the AAE $W_\varphi$ and PAE $W_\theta$ on scanning speed $\upsilon$ as a main parameter of the treatment of surface by NLL was studied. It was found that there is a certain correlation between both AAE $W_\varphi$ and PAE $W_\theta$ for the same experimental conditions (*e.g.* scanning speed $\upsilon$ during the process of nano-structuring of Ti layer by NLL). It can be concluded that the choice of scanning speed during the process of Ti surface nano-structuring allows changing values in both PAE $W_\theta$ and AAE $W_\varphi$. This is closely related to the depth of nano-grooves obtained after the process of surfaces nano-strucruring and their further modification by additional layers (*e. g.* ITO, polyimide PI2555). It was found that the certain correlation between the contact angle *β* of nematic droplet deposited on the FTAL/STAL and scanning speed *υ* during the process of NLL and therefore between the contact angle *β* and PAE $W_\varphi$ and AAE $W_\theta$ take place.

Acknowledgements

The authors thank W. Becker (Merck, Darmstadt, Germany) for his generous gift of nematic liquid crystals E7 and MLC-6609, A. Kratz (Merck, Germany) for her produce of the Licristal brochure and field service specialist IV V. Danylyuk (Dish LLC, USA) for his gift of some Laboratory equipments. The authors thank Dr. T. Kosa (AlphaMicron, Inc, Kent OH, USA) for his explanation of some experimental nuances and Dr. S. Lukyanets (Institute of Physics, NAS of Ukraine) for the helpful discussions.

Caption

Figure 1. Schematic presentation of director $\vec{n}$ in the bulk of LC, which is bordered on the nano-structured aligning layer.

Figure 2. Schematic image of the nano-structured Ti layer treated by means of NLL, with area 5×5 mm$^2$ and additionally coated with ITO layer (FTAL).

Figure 3. Optical scheme of Senarmont compensator, consisting of: He-Ne laser LGN-111 ($\lambda$ = 632 nm and power P = 15 mW); P and A – polarizer and analyzer (Glan-Thompson prizms); Sample – symmetrical LC cell, consisting of a pair of substrates, possessing nano-structured Ti layers by means of NLL, additionally coated with ITO (FTAL), or FTAL additionally coated with polyimide PI2555 (STAL); $\lambda/4$ – quarter-wave plate (QWP); V - low frequency signal generator GZ-109; PD – silicon photodetector FD-18K, connected to the oscilloscope Hewlett Packard 54602B.

Figure 4. (a) The scheme to measure of the contact angle β, consisting of horizontal microscope with an ocular micrometer in an eyepiece, the studying aligning layer and deposited droplet (E7, MLC-6609 or glycerol). (b) The photography of capillary with geometrical dimensions, used for the deposition of droplets. The capillaries used

possessed the outer diameter $d_{out}$ = 350 μm and inner opening with $d_{in}$ = 150 μm. (c). The photography of the glycerol droplet with certain diameter D and height H. The droplet was deposited on the pure nano-structured Ti layer (treated with scanning speed $\upsilon$ = 2000 mm/s) and observed through an eyepiece of horizontal microscope.

Figure 5. AFM images of the (a) pure nano-sturctured Ti layer, (b) pure nano-structuted Ti layer coated with the ITO film – FTAL and (c) pure nano-structured Ti layer covered the ITO film (FTAL) coated with the polyimide PI2555 film - STAL. The cross section of nano-grooves of the layer: (d) the pure nano-structured Ti, possessing average values of the period $\Lambda$ = 910 nm and depth A = 270 nm (solid red squares), (e) the FTAL, having average values of the period $\Lambda$ = 890 nm and depth A = 291 nm (solid blue circles) and (f) the STAL, possessing average values of the period $\Lambda$ = 890 nm and depth A = 256 nm (solid green triangles). Parameters of the NLL are: the scanning speed $\upsilon$ = 1500 mm/s and the laser pulse fluence $J$ = 0.55 J/cm$^2$.

Figure 6. Dependence of the depth of nano-grooves A on the scanning speed $\upsilon$ for: 1) the pure nano-structured Ti layer (curve1, opened red squares); 2) the FTAL, consisting of the pure nano-structured Ti layer additionally coated with ITO (curve 2, opened blue squares); 3) the STAL, consisting of the FTAL additionally coated with polyimide PI2555 film (curve3, opened green treangles). During of the NLL technique of the Ti surface the LPF $J$ = 0.55 J/cm$^2$. The dashed curve is a guide to the eye.

Figure 7. Dependence of AE $W_B$ on (a) the scanning speed $\upsilon$ and (b) depth of nano-grooves A for: 1) the pure nano-structured Ti layer (curve 1); 2) the FTAL, consisting of pure nano-structured Ti layer additionally coated with ITO (curve 2); 3) the STAL,

consisting of the FTAL additionally coated with polyimide PI2555 film (curve 3). The value of LPF during of the NLL technique of the Ti surface was $J = 0.55$ J/cm$^2$. For calculation of AE $W_B$ the Frank constants of the nematic E7 were used. The dashed line is a guide to the eye.

Figure 8. Dependencies of (a) twist angle $\varphi$ and (b) calculated AAE $W_\varphi$ of the LC cells, filled by nematic E7 ($\Delta\varepsilon > 0$), on scanning speed $v$ for the pure nano-structured Ti layer (curves 1), the FTAL, consisting of the pure nano-structured Ti layer additionally coated with ITO (curves 2) and the STAL, consisting of the FTAL additionally coated with polyimide PI2555 (curves 3). Data from all curves (opened symbols) were measured in 30 min after filling of LC cells. The dashed line is a guide to the eye.

Figure 9. Dependence of the AAE $W_\varphi$ on: (a) the depth of nano-grooves A (solid symbols) and (b) the twist angle $\varphi$ (opened symbols) for various aligning layers: 1 – the pure nano-structured Ti layer, 2 – the FTAL and 3 – the STAL. The tilt angle $\psi$ of the linear fit 1, 2 and 3 is about 16, 18 and 29 degrees, respectively. LC cells were filled by the nematic E7 ($\Delta\varepsilon > 0$).

Figure 10. Dependence of (a) the twist angle $\varphi$ and (b) the calculated AAE $W_\varphi$ of the LC cells, filled by the nematic MLC-6609 ($\Delta\varepsilon > 0$), on the scanning speed $v$ for the FTAL (curves 1), consisting of the pure nano-structured Ti layer additionally coated with ITO, and the STAL (curves 2), consisting of the FTAL additionally coated with the polyimide PI2555. The dashed line is a guide to the eye.

Figure 11. Dependence of the pretilt angle $\theta_p$ of the symmetrical LC cell on the scanning speed $v$ for: (a) the unstable planar alignment of the nematic E7 (curve 1-0, solid blue squares) and MLC-6609 (curve 2-0, solid red circles) for the FTAL. The stable planar alignment of the nematic E7 (curve 3, opened green triangles) and MLC-6609 (curve 4, solid black diamonds) for the STAL; (b) the P-H transition of the nematic E7 and MLC-6609 for the FTAL. The curve 1-0 (solid blue squares) and the curve 2-0 (solid red circles) - the pretilt of the planar alignment of the nematic E7 and MLC-6609, respectively. The curve 1 (opened blue squares) and the curve 2 (opened red circles) - the pretilt homeotropic alignment of the nematic E7 and MLC-6609, respectively. The inset depicts the $\theta_p(v)$ for the FTAL after the P-H transition in a large scale. The dashed line is a guide to the eye.

Figure 12. Dependence of the optical phase retardation function $R(V-V^*)$ on $(V-V^*)$ of the LC cell consisting with: (a) FTAL and filling by the nematic MLC 6609 ($\Delta\varepsilon < 0$) and (b) STAL and filling by the nematic E7 ($\Delta\varepsilon > 0$). For both types of the aligning layers, the scanning speed $v$ and LPF $J$ were 500 mm/s and 0.55 J/cm$^2$, respectively. The thickness of LC cell was 23.5 μm (a) and 25.7 μm (b). The values of the linear fit parameters are: (a) A = 11.8649, B = - 0.5512, $V_{min}$ = 9.7V and $V_{max}$ = 17.9V, (b) A = 0.6628, B = 0.0093, $V_{min}$ = 2.3V and $V_{max}$ = 5.7V. Estimated values of the PAE $W_\theta$ for (a) FTAL and (b) STAL are about $0.7\times10^{-4}$ and $0.2\times10^{-4}$ J/m$^2$, respectively.

Figure 13. Dependencies of the calculated PAE $W_\theta$ on: (a) the scanning speed $v$ and (b) pretilt angle $\theta_p$. LC cell consists of two aligning layers: FTAL (curve 1, solid blue squares) or STAL (curve 2, solid green triangles). LC cells were filled by the nematic MLC-6609 (curve 1) and E7 (curve 2), respectively. The slopes $\phi$ for the FTAL (line 1,

opened blue squares) and the STAL (line 2, opened green triangles) are 86 and 78 degrees, respectively. The dashed line is a guide to the eye.

Figure 14. The correlation between PAE $W_\theta$ and AAE $W_\varphi$ for: FTAL (opened blue diamonds) and STAL (solid green spheres). LC cells were filled by the nematic MLC-6609 (line 1) and E7 (line 2).

Figure 15. Dependence of the contact angle $\beta$ on the number of rubbing $N_{rubb}$ for: (a) polyimide ODAPI layer for droplet of glycerol (curves 1, opened red squares), nematic E7 (curves 2, opened blue circles) and MLC-6609 (curves 3, opened green triangles), measured along direction of rubbing and; (b) polyimide PI2555 layer for droplet of nematic E7, measured along (curve 1, opened blue diamonds) and across (curve 2, solid blue diamonds) direction of rubbing. The dashed curves are guide to the eye.

Figure 16. Photography of droplet of glycerol and two nematics, namely MLC-6609 and E7, placed at the aligning surface FTAL, obtained under NLL treatment of the Ti layer with the laser pulse fluence $J = 0.55$ J/cm$^2$ and the scanning speed $\upsilon = 2000$ mm/s with further modification by means of the ITO-coating. FTAL possesses by nano-grooves with average values of the period $\Lambda = 890$ nm and depth A = 291 nm. The value of the AAE $W_\varphi$ was about $0.47 \times 10^{-4}$ J/m$^2$.

Figure 17. Dependence of the contact angle $\beta$ of droplet of the glycerol (curves 1), nematic E7 (curves 2) and nematic MLC-6609 (curves 3) deposited on (a) FTAL and (b) STAL, on scanning speed $\upsilon$ during process of NLL. Contact angles $\beta$ were measured along the rubbing direction. The dashed curve is guide to the eye.

Figure 18. Dependence of the calculated values of the AAE $W_\varphi$ (lines 1) and the PAE $W_\theta$ (lines 2) on contact angle $\beta$ of the droplet of: (a) nematic MLC-6609 deposited on FTAL and (b) nematic E7 deposited on STAL. Contact angles $\beta$ were measured along the rubbing direction. The dashed line is a guide to the eye.